\documentclass{article}
\usepackage{graphicx}
\begin{document}
\title{QUARK-ANTIQUARK PAIRING INDUCED BY INSTANTONS IN QCD}
\author{Nguyen Van Hieu, Nguyen Hung Son\\
Ngo Van Thanh and Hoang Ba Thang\\
{\small Institute of Physics, P. O. Box 429, Boho, Hanoi 10000, Vietnam}}
\date{}
\maketitle
\begin{abstract}
The quark-antiquark pairing due to the four-fermion direct coupling induced
by the instantons in QCD is studied in the framework of the functional
integral method. The integral equations for the order parameters are
derived. Several classes of the solutions of these non-linear equations are
considered in details. Their implications for the dynamical symmetry
breakings are discussed.
\end{abstract}

\section{INTRODUCTION.}

\qquad The superconducting pairing of quarks due to the gluon exchange in
QCD with the formation of the diquark condensate was proposed by Barrois$%
^{\left[ 1\right] }$ and Frautschi$^{\left[ 2\right] }$ since more than two
decades and then studied by Bailin and Love$^{\left[ 3\right] }$, Donoglue
and Sateesh$^{\left[ 4\right] }$, Iwasaki and Iwado$^{\left[ 5\right] }$.
Recently in a series papers by Alford, Rajagopal and Wilczek$^{\left[
6\right] }$, Sch\"{a}fer and Wilczek$^{\left[ 7\right] }$, Rapp,
Sch\"{a}fer, Shuryak and Velkovsky$^{\left[ 8\right] }$, Evans, Hsu and
Schwetz$^{\left[ 9\right] }$, Hieu and Tuong$^{\left[ 10\right] }$, Son$%
^{\left[ 11\right] }$, and others there arose a new interest to the
existence of the diquark Bose condensate in the QCD dense matter - the color
superconductivity. The connection between the color superconductivity and
the chiral phase transition in\ QCD was studied by Berges and Rajagopal$%
^{\left[ 12\right] }$, Harada and Shibata$^{\left[ 13\right] }$. There
exists also the spontaneous parity violation, as it was shown by Pisarski
and Rischke$^{\left[ 14\right] }$.

The present work is devoted to the study of the quark-antiquark pairing due
to some direct four-fermion coupling in the framework of the functional
integral approach$^{\left[ 15\right] }$. For the systems of quarks and
antiquarks with $N_{c}$ colors and $N_{f}=2$ flavors this direct
four-fermion coupling might be induced by the instantons$^{\left[
8,16,17\right] }$. We derive the general integral equations for the order
parameters of the Bose condensate of the quark-antiquark pairs in the
interacting systems with any direct four-fermion coupling. By establishing
the existence of the order parameters with definite symmetry properties in
the system with $N_{c}$ colors and two flavors we come to the
conclusions on the dynamical spontaneous breaking of the corresponding
symmetries.

\section{BOSONIC COMPOSITE FIELD.}

\qquad Denote $\psi _{A}\left( x\right) $ the quark field, where $A=\left(
\alpha ai\right) $ is the set consisting of the Dirac spinor index $\alpha
= 1,2,3,4$ and the color and flavor indices $a=1,2,...N_{c}$ and $
i=1,2,...N_{f}$. The internal symmetry groups are assumed to be $SU\left(
N_{c}\right) _{c}$ and $SU\left( N_{f}\right) _{f}$. We work in the
imaginary time formalism and denote $x=\left( \tau ,\mathbf{x}\right) $ a
vector in the Euclidean four-dimensional space.

The partition function of the system of free quarks and antiquarks at the
temperature $T$ can be expressed in the form of the functional integral 
\begin{equation}
Z_{0}=\int \left[ D\psi \right] \left[ D\overline{\psi }\right] \exp \left\{
-\int dx\overline{\psi }^{A}\left( x\right) D_{A}^{B}\psi _{B}\left(
x\right) \right\}  \label{1}
\end{equation}
with the brief notations 
\begin{eqnarray}
\int dx &=&\int_{0}^{\beta }d\tau \int d\mathbf{x},\,\,\,\,\,\,\,\,\,\,\,\,%
\,\,\;\quad \beta =\frac{1}{kT}\,\,\,,  \nonumber \\
D_{A}^{B} &=&\left[ \gamma _{4}\left( \frac{\partial }{\partial \tau }-\mu
\right) +\mathbf{\gamma \nabla }+M\right] _{A}^{B}\,,  \label{2}
\end{eqnarray}
where $k\,$is the Boltzmann constant, $\mu $ is the chemical potential and $%
M$ is the bare quark mass. We write the interaction Lagrangian of
the four-fermion coupling in the most general form 
\begin{equation}
L_{int}=\frac{1}{2}\overline{\psi }^{A}\left( x\right) \psi _{B}\left(
x\right) U_{AC\,}^{BD}\overline{\psi }^{C}\left( x\right) \psi _{D}\left(
x\right) .  \label{3}
\end{equation}
The partition function of the interacting system equals 
\begin{eqnarray}
Z &=&\int \left[ D\psi \right] \left[ D\overline{\psi }\right] \exp \left\{
-\int dx\overline{\psi }^{A}\left( x\right) D_{A}^{B}\psi _{B}\left(
x\right) \right\}  \nonumber \\
&&\exp \left\{ \frac{1}{2}\int dx\overline{\psi }^{A}\left( x\right) \psi
_{B}\left( x\right) U_{AC\,}^{BD}\overline{\psi }^{C}\left( x\right) \psi
_{D}\left( x\right) \right\} .  \label{4}
\end{eqnarray}

Following the method in the functional integral approach$^{\left[ 15\right]
} $ we introduce some bosonic composite field $\Phi _{B}^{A}\left( x\right) $
and the functional integral 
\begin{equation}
Z_{0}^{\Phi }=\int \left[ D\Phi \right] \exp \left\{ -\frac{1}{2}\int dx\Phi
_{B}^{A}\left( x\right) U_{AC\,}^{BD}\Phi _{D}^{C}\left( x\right) \right\} .
\label{5}
\end{equation}
We have the Hubbard-Stratonovich transformation 
\begin{eqnarray}
&&\exp \left\{ \frac{1}{2}\int dx\overline{\psi }^{A}\left( x\right) \psi
_{B}\left( x\right) U_{AC\,}^{BD}\overline{\psi }^{C}\left( x\right) \psi
_{D}\left( x\right) \right\} =\frac{1}{Z_{0}^{\Phi }}\int \left[ D\Phi
\right]  \nonumber \\
&&\hspace{1cm}\exp \left\{ -\frac{1}{2}\int dx\Phi _{B}^{A}\left( x\right)
U_{AC\,}^{BD}\Phi _{D}^{C}\left( x\right) \right\} \exp \left\{ -\int dx%
\overline{\psi }^{A}\left( x\right) \psi _{B}\left( x\right) \Delta
_{A}^{B}\left( x\right) \right\} ,  \label{6}
\end{eqnarray}
where 
\begin{equation}
\Delta _{A}^{B}\left( x\right) =U_{AC\,}^{BD}\Phi _{D}^{C}\left( x\right) .
\label{7}
\end{equation}
\hspace{-0.05cm}Denote $S_{A}^{B}\left( x-y\right) $ the two-point Green
function of the free quark field, 
\begin{equation}
D_{A}^{B}S_{B}^{C}\left( x-y\right) =\delta _{A}^{C}\delta \left( x-y\right)
.  \label{8}
\end{equation}
Substituting the expression (6) into the r.h.s. of the formula (4) and
performing the functional integration over the fermionic variables, we
obtain the partition function $Z\,$in the form of a functional integral over
the bosonic integration variables 
\begin{equation}
Z=\frac{Z_{0}}{Z_{0}^{\Phi }}\int \left[ D\Phi \right] \exp \left\{ I_{\mathrm{eff}}
\left[ \Phi \right] \right\}  \label{9}
\end{equation}
with the effective action of the bosonic composite field 
\begin{equation}
I_{\mathrm{eff}}\left[ \Phi \right] =-\frac{1}{2}\int dx\Phi _{B}^{A}\left(
x\right) U_{AC\,}^{BD}\Phi _{D}^{C}\left( x\right) +W\left[ \Delta \right] ,
\label{10}
\end{equation}
\begin{equation}
W\left[ \Delta \right] =\sum_{n=1}^{\infty }W^{\left( n\right) }\left[
\Delta \right] ,  \label{11}
\end{equation}
\[
W^{\left( 1\right) }\left[ \Delta \right] =\int dx\Delta _{A}^{B}\left(
x\right) S_{B}^{A}\left( 0\right) , 
\]
\begin{equation}
W^{\left( 2\right) }\left[ \Delta \right] =-\frac{1}{2}\int dx_{1}\int
dx_{2}\Delta _{A_{1}}^{B_{1}}\left( x_{1}\right) S_{B_{1}}^{A_{2}}\left(
x_{1}-x_{2}\right) \Delta _{A_{2}}^{B_{2}}\left( x_{2}\right)
S_{B_{2}}^{A_{1}}\left( x_{2}-x_{1}\right) ,  \label{12}
\end{equation}
\begin{eqnarray*}
W^{\left( 3\right) }\left[ \Delta \right] &=&\frac{1}{3}\int dx_{1}\int
dx_{2}\int dx_{3} \\
&&\Delta _{A_{1}}^{B_{1}}\left( x_{1}\right) S_{B_{1}}^{A_{2}}\left(
x_{1}-x_{2}\right) \Delta _{A_{2}}^{B_{2}}\left( x_{2}\right)
S_{B_{2}}^{A_{3}}\left( x_{2}-x_{3}\right) \Delta _{A_{3}}^{B_{3}}\left(
x_{3}\right) S_{B_{3}}^{A_{1}}\left( x_{3}-x_{1}\right) ,
\end{eqnarray*}
\begin{eqnarray*}
W^{\left( n\right) }\left[ \Delta \right] &=&\frac{\left( -1\right) ^{n+1}}{n%
}\int dx_{1}...\int dx_{n}\Delta _{A_{1}}^{B_{1}}\left( x_{1}\right)
S_{B_{1}}^{A_{2}}\left( x_{1}-x_{2}\right) \\
&&\Delta _{A_{2}}^{B_{2}}\left( x_{2}\right) ....S_{B_{n-1}}^{A_{n}}\left(
x_{n-1}-x_{n}\right) \Delta _{A_{n}}^{B_{n}}\left( x_{n}\right)
S_{B_{n}}^{A_{1}}\left( x_{n}-x_{1}\right) \\
&&.......
\end{eqnarray*}
From this effective action we derive the field equation 
\begin{equation}
\Delta _{C}^{D}\left( x\right) =U_{CA\,}^{DB}G_{B}^{A}\left( x,x\right) ,
\label{13}
\end{equation}
where the two-point Green function of the quark field in the presence of the
condensate $G_{B}^{A}\left( y,x\right) $ is determined by the
Schwinger-Dyson equation 
\begin{equation}
G_{B}^{A}\left( y,x\right) =S_{B}^{A}\left( y-x\right) -\int
dzS_{B}^{C}\left( y-z\right) \Delta _{C}^{D}\left( z\right) G_{D}^{A}\left(
z,x\right) .  \label{14}
\end{equation}

\section{GENERAL EQUATIONS AND RELATIONS FOR ORDER PARAMETERS.}

\qquad Now we study the constant solution 
\begin{equation}
\Delta _{B}^{A}\left( x\right) =\Delta _{\,\,B}^{0A}=\mathrm{const}.
\label{15}
\end{equation}
of the system of equations (13) and (14). The non-vanishing components of
the constant rank 2 spinor (15) can be considered as the order parameters of
the quark-antiquark pair Bose condensate. For the special class (15) of the
bosonic field $\Delta _{B}^{A}\left( x\right) $ the Green functions $%
G_{B}^{A}\left( y,x\right) $depend only on the coordinate difference 
\[
G_{B}^{A}\left( y,x\right) =G_{B}^{A}\left( y-x\right) . 
\]
Denoting $\widetilde{G}_{B}^{A}\left( \mathbf{p},\varepsilon _{n}\right) $%
and $\widetilde{S}_{B}^{A}\left( \mathbf{p},\varepsilon _{n}\right) $ the
Fourier transforms of $G_{B}^{A}\left( y-x\right) \,$and $S_{B}^{A}\left(
y-x\right) $ resp., 
\[
\varepsilon _{n}=\left( 2n+1\right) \frac{\pi }{\beta }, 
\]
$n$ being integers, we rewrite the equations (13) and (14) in the form 
\begin{eqnarray}
\Delta _{C}^{0D} &=&U_{CA}^{DB}\frac{1}{\beta }\sum_{n}\frac{1}{\left( 2\pi
\right) ^{3}}\int d\mathbf{p}\widetilde{G}_{B}^{A}\left( \mathbf{p}%
,\varepsilon _{n}\right)  \label{16} \\
\widetilde{G}_{B}^{A}\left( \mathbf{p},\varepsilon _{n}\right) &=&\widetilde{%
S}_{B}^{A}\left( \mathbf{p},\varepsilon _{n}\right) -\widetilde{S}%
_{B}^{C}\left( \mathbf{p},\varepsilon _{n}\right) \Delta _{C}^{0D}\widetilde{%
G}_{D}^{A}\left( \mathbf{p},\varepsilon _{n}\right)  \label{17}
\end{eqnarray}
Introducing the matrices $\widehat{\Delta }^{0},\widehat{G}\left( \mathbf{p}%
,\varepsilon _{n}\right) $ and $\widehat{S}\left( \mathbf{p},\varepsilon
_{n}\right) $ with the elements $\Delta _{B}^{0\,A},\widetilde{G}%
_{B}^{A}\left( \mathbf{p},\varepsilon _{n}\right) $and $\widetilde{S}%
_{B}^{A}\left( \mathbf{p},\varepsilon _{n}\right) ,$ and the inverse matrix 
\begin{equation}
\frac{1}{\widehat{S}\left( \mathbf{p},\varepsilon _{n}\right) }=i\left[
\gamma _{4}\left( \varepsilon _{n}+i\mu \right) +\mathbf{\gamma p}\right]
+M=i\widehat{p}+M,  \label{18}
\end{equation}
from the Schwinger-Dyson equation (17) we obtain 
\begin{equation}
\frac{1}{\widehat{G}\left( \mathbf{p},\varepsilon _{n}\right) }=\frac{1}{%
\widehat{S}\left( \mathbf{p},\varepsilon _{n}\right) }+\widehat{\Delta }^{0}.
\label{19}
\end{equation}
Using the expressions (10)-(12) of the partition function and the field
equations (13) and (14) we can derive also the expression of the free energy
density of the system. In the case of the constant field (15) we have$%
^{\left[ 15\right] }$ 
\begin{equation}
F\left[ \widehat{\Delta }^{0}\right] =-\frac{1}{\beta ^{2}}\sum_{n}\frac{1}{%
\left( 2\pi \right) ^{3}}\int d\mathbf{p}\mathrm{Tr}\left\{ \widehat{\Delta }%
^{0}\left[ \int_{0}^{1}\widehat{G}^{\omega }\left( \mathbf{p},\varepsilon
_{n}\right) d\omega -\frac{1}{2}\widehat{G}\left( \mathbf{p},\varepsilon
_{n}\right) \right] \right\} ,  \label{20}
\end{equation}
where $\widehat{G}\left( \mathbf{p},\varepsilon _{n}\right) $was given by
the equation (19) and $\widehat{G}^{\omega }\left( \mathbf{p},\varepsilon
_{n}\right) $ is determined by a similar one with the replacement of $%
\widehat{\Delta }^{0}$ by $\omega \widehat{\Delta }^{0}$:

\begin{equation}
\frac{1}{\widehat{G}^{\omega }\left( \mathbf{p},\varepsilon _{n}\right) }=%
\frac{1}{\widehat{S}\left( \mathbf{p},\varepsilon _{n}\right) }+\omega 
\widehat{\Delta }^{0}.  \label{21}
\end{equation}

From the rotational invariance it follows that the rank 2 spinor $\Delta
_{\left( \beta bj\right) }^{0\left( \alpha ai\right) }$ has the most general
form 
\begin{equation}
\Delta _{\left( \beta bj\right) }^{0\left( \alpha ai\right) }=\delta _{\beta
}^{\alpha }\Sigma \,_{\left( bj\right) }^{\left( ai\right) }+\left( \gamma
_{4}\right) _{\beta }^{\alpha }\Sigma \,_{\left( bj\right) }^{^{\prime
}\,\,\left( ai\right) }+i\left( \gamma _{5}\right) _{\beta }^{\alpha }\Pi
\,_{\left( bj\right) }^{\left( ai\right) }+i\left( \gamma _{4}\gamma
_{5}\right) \Pi \,_{\left( bj\right) }^{\prime \,\,\left( ai\right) }.
\label{22}
\end{equation}
In order to avoid the lengthy calculations we consider the systems with the
order parameters having a special form 
\begin{equation}
\Delta _{\left( \beta bj\right) }^{0\left( \alpha ai\right) }=\delta _{\beta
}^{\alpha }\Sigma \,_{\left( bj\right) }^{\left( ai\right) }\,+i\left(
\gamma _{5}\right) _{\beta }^{\alpha }\Pi \,_{\left( bj\right) }^{\left(
ai\right) }.  \label{23}
\end{equation}
There is some correspondence between the existence of the non-vanishing
components of $\Delta _{\,\left( \beta bj\right) }^{0\left( \alpha ai\right)
}$ and the dynamical spontaneous symmetry breakings in these systems:

\begin{itemize}
\item[a)]  Spontaneous parity violation $\leftrightarrow \Pi _{\left(
bj\right) }^{\left( ai\right) }\neq 0.$

\item[b)]  Spontaneous color symmetry breaking $\leftrightarrow \Sigma
\,_{\left( bj\right) }^{\left( ai\right) }\neq \delta _{b}^{a}$ $\Sigma
\,_{j}^{i}\,\,$and/or $\Pi \,_{\left( bj\right) }^{\left( ai\right) }$ $\neq
\delta _{b}^{a}\Pi \,_{j}^{i}.$

\item[c)]  Spontaneous flavor symmetry breaking $\leftrightarrow \Sigma
\,_{\left( bj\right) }^{\left( ai\right) }\neq \delta _{j}^{i}\Sigma
\,_{b}^{a}\,\,$and/or $\Pi \,_{\left( bj\right) }^{\left( ai\right) }\neq
\delta \,_{j}^{i}\Pi _{b}^{a}.$

\item[d)]  Spontaneous chiral symmetry breaking $\leftrightarrow M=0,\Sigma
\,_{\left( bj\right) }^{\left( ai\right) }\,\neq 0\,$ and/or $\Pi \,_{\left(
bj\right) }^{\left( ai\right) }\neq 0.$
\end{itemize}

In the study of the QCD vacuum one often used the physical quantity

\[
n=\left\langle \overline{\psi }^{A}\left( x\right) \psi _{A}\left( x\right)
\right\rangle 
\]
which was called the quark condensate. This constant is expressed in terms
of the covariant components of the order parameters $\Sigma \,_{\left(
bj\right) }^{\left( ai\right) }$ and/or $\Pi \,_{\left( bj\right) }^{\left(
ai\right) }$.

For the instanton induced four-fermion direct coupling in the system with $%
N_{c}$ colors and two flavors we have$^{\left[ 8,16,17\right] }$ 
\begin{eqnarray}
U_{AC}^{BD} &=&\frac{1}{2}\left\{ g_{1}\left[ \delta _{\alpha }^{\beta
}\delta _{\gamma }^{\delta }+\left( \gamma _{5}\right) _{\alpha }^{\beta
}\left( \gamma _{5}\right) _{\gamma }^{\delta }\right] +g_{2}\left( \sigma
_{\mu \nu }\right) _{\alpha }^{\beta }\left( \sigma _{\mu \nu }\right)
_{\gamma }^{\delta }\right\} \delta _{a}^{b}\delta _{c}^{d}\left[ \delta
_{i}^{j}\delta _{k}^{\ell }-\left( \mathbf{\tau }\right) _{i}^{j}\left( 
\mathbf{\tau }\right) _{k}^{\ell }\right] \varphi  \nonumber  \label{24} \\
&&-\frac{1}{2}\left\{ g_{1}\left[ \delta _{\alpha }^{\delta }\delta _{\gamma
}^{\beta }+\left( \gamma _{5}\right) _{\alpha }^{\delta }\left( \gamma
_{5}\right) _{\gamma }^{\beta }\right] +g_{2}\left( \sigma _{\mu \nu
}\right) _{\alpha }^{\delta }\left( \sigma _{\mu \nu }\right) _{\gamma
}^{\beta }\right\} \delta _{a}^{d}\delta _{c}^{b}\left[ \delta _{i}^{\ell
}\delta _{k}^{j}-\left( \mathbf{\tau }\right) _{i}^{\ell }\left( \mathbf{%
\tau }\right) _{k}^{j}\right] \varphi ,\qquad
\end{eqnarray}
where $\varphi $ is the form-factor. \noindent With the positive constant $%
g_{1}$ the quark-antiquark effective interaction is attractive in the
pseudoscalar isovector (pion) and scalar isoscalar ($\sigma $ - meson)
channels$^{\left[ 18\right] }$. Substituting the expressions (23) and (24)
into the r.h.s. of the equations (16) and (17) and introducing the $%
N_{c}N_{f}\times N_{c}N_{f}$ matrices $\widehat{\Sigma }$ and $\widehat{\Pi }
$ with the elements $\Sigma \,_{\left( bj\right) }^{\left( ai\right) }$ and $%
\Pi \,_{\left( bj\right) }^{\left( ai\right) }$, we obtain the system of
equations for these order parameters in the matrix form: 
\begin{eqnarray}
\Sigma \,_{\left( ck\right) }^{\left( d\ell \right) } &=&V\,_{\left(
ck\right) \left( ai\right) }^{\left( d\ell \right) \left( bj\right) }\frac{1%
}{\beta }\sum_{n}\frac{1}{2\pi ^{2}}\int \varphi \left( p\right)  \nonumber
\\
&&\left[ \left( M+\widehat{\Sigma }\right) \frac{1}{\left( \varepsilon
_{n}+i\mu \right) ^{2}+p^{2}+\left( M+\widehat{\Sigma }\right) ^{2}+\widehat{%
\Pi }^{2}}\right] _{bj}^{ai}p^{2}dp,  \label{25}
\end{eqnarray}
\begin{eqnarray}
\Pi \,_{\left( ck\right) }^{\left( d\ell \right) } &=&-V\,_{\left( ck\right)
\left( ai\right) }^{\left( d\ell \right) \left( bj\right) }\frac{1}{\beta }%
\sum_{n}\frac{1}{2\pi ^{2}}\int \varphi \left( p\right)  \nonumber \\
&&\left[ \widehat{\Pi }\frac{1}{\left( \varepsilon _{n}+i\mu \right)
^{2}+p^{2}+\left( M+\widehat{\Sigma }\right) ^{2}+\widehat{\Pi }^{2}}\right]
_{bj}^{ai}p^{2}dp,  \label{26}
\end{eqnarray}
\begin{equation}
V\,_{\left( ck\right) \left( ai\right) }^{\left( d\ell \right) \left(
bj\right) }=2g_{1}\delta _{a}^{b}\delta _{c}^{d}\left[ \delta _{i}^{j}\delta
_{k}^{\ell }-\left( \mathbf{\tau }\right) _{i}^{j}\left( \mathbf{\tau }%
\right) _{k}^{\ell }\right] -\left( g_{1}+6g_{2}\right) \delta
_{a}^{d}\delta _{c}^{b}\left[ \delta _{i}^{\ell }\delta _{k}^{j}-\left( 
\mathbf{\tau }\right) _{i}^{\ell }\left( \mathbf{\tau }\right)
_{k}^{j}\right] .  \label{27}
\end{equation}

\section{ORDER PARAMETERS WITH DEFINITE SYMMETRY PROPERTIES.}

\qquad Let us study the solution of the system of equations (25) and (26)
for the order parameters having definite transformation properties with
respect to different symmetry groups. We use following notations: 
\begin{eqnarray}
K_{\Sigma }^{\Pi }\left( M,\beta ,\mu \right) &=&\frac{1}{8\pi ^{2}}\int 
\frac{\varphi \left( p\right) }{E_{\Sigma }^{\Pi }\left( p,M\right) }\left\{ 
\mathrm{th}\frac{\beta \left[ E_{\Sigma }^{\Pi }\left( p,M\right) +\mu \right] 
}{2}\right.  \nonumber \\
&&+\left. \mathrm{th}\frac{\beta \left[ E_{\Sigma }^{\Pi }\left( p,M\right)
-\mu \right] }{2}\right\} p^{2}dp,  \label{28}
\end{eqnarray}
\begin{equation}
E_{\Sigma }^{\Pi }\left( p,M\right) =\left[ p^{2}+\left( M+\Sigma \right)
^{2}+\Pi ^{2}\right] ^{1/2}.  \label{29}
\end{equation}

\begin{equation}
g=2\left[ \left( 2N_{c}+1\right) g_{1}+6g_{2}\right] .  \label{30}
\end{equation}
In the case of the instanton induced effective four-fermion coupling the
constant $g$ is positive. In order to have a more complete presentation as
well as to provide a comparison we consider also the case $g<0$. For
studying the spontaneous breaking of the parity conservation or/and the
flavor symmetry without that of the color symmetry we write the order
parameters in the form

\begin{equation}
\Sigma _{\left( bj\right) }^{\left( ai\right) }=\delta _{b}^{a}\left[ \delta
_{j}^{i}\Sigma _{s}+\left( \mathbf{\tau n}\right) _{j}^{i}\Sigma _{t}\right]
,  \label{31}
\end{equation}

\begin{equation}
\Pi _{\left( bj\right) }^{\left( ai\right) }=\delta _{b}^{a}\left[ \delta
_{j}^{i}\Pi _{s}+\left( \mathbf{\tau n}\right) _{j}^{i}\Pi _{t}\right] ,
\label{32}
\end{equation}
where $\mathbf{n}$ is some three-dimensional unit vector, and set

\begin{equation}
\Sigma ^{\left( \pm \right) }=\Sigma _{s}\pm \Sigma _{t},  \label{33}
\end{equation}

\begin{equation}
\Pi ^{\left( \pm \right) }=\Pi _{s}\pm \Pi _{t}.  \label{34}
\end{equation}
As the simple examples we consider in details three following particular
phases:

\begin{description}
\item[1)]  $\qquad \Pi _{j}^{i}=0;$

\item[2)]  $\qquad \Pi _{s}\neq 0,\qquad \Pi _{t}=0;$

\item[3)]  $\qquad \Pi _{t}\neq 0,\qquad \Sigma _{t}=0.$
\end{description}

\noindent \underline{Phase 1}. In this phase we have the system without the
spontaneous parity violation. From the equations (25) for $\Sigma _{\left(
bj\right) }^{\left( ai\right) }$ we derive a system of two equations for $%
\Sigma ^{\left( \pm \right) }$:

\begin{eqnarray}
\Sigma ^{\left( +\right) } &=&\left[ M+\Sigma ^{\left( -\right) }\right]
gK_{\Sigma ^{\left( -\right) }}^{0}\left( M,\beta ,\mu \right) ,  \nonumber
\\
\Sigma ^{\left( -\right) } &=&\left[ M+\Sigma ^{\left( +\right) }\right]
gK_{\Sigma ^{\left( +\right) }}^{0}\left( M,\beta ,\mu \right) .  \label{35}
\end{eqnarray}
In the case of the non-vanishing bare quark mass,

\[
M\neq 0, 
\]
from these equations it follows that

\begin{equation}
\Sigma ^{\left( +\right) }=\Sigma ^{\left( -\right) },  \label{36}
\end{equation}
and therefore

\begin{equation}
\Sigma _{t}=0.  \label{37}
\end{equation}
This means that the flavor symmetry also cannot be spontaneously broken. The
constant $\Sigma _{s}$ is determined by the equation

\begin{equation}
\frac{\Sigma _{s}}{M}=\frac{gK_{\Sigma _{s}}^{0}\left( M,\beta ,\mu \right) 
}{1-gK_{\Sigma _{s}}^{0}\left( M,\beta ,\mu \right) }=\frac{gK_{0}^{0}\left(
M+\Sigma _{s},\beta ,\mu \right) }{1-gK_{0}^{0}\left( M+\Sigma _{s},\beta
,\mu \right) }.  \label{38}
\end{equation}
Its solution always exists. The values of the integral $gK_{\Sigma
}^{0}\left( M,\beta ,\mu \right) $at different sets of the values of the
physical parameters of the system are given in the Appendix. The
quasiparticle with the momentum $\mathbf{p}$ has the energy $E_{\Sigma
_{s}}^{0}\left( \mathbf{p},M\right) $ determined by the formula (29), 
\begin{equation}
E_{\Sigma _{s}}^{0}\left( \mathbf{p},M\right) =\left[ \mathbf{p}^{2}+\left(
M+\Sigma _{s}\right) ^{2}\right] ^{1/2}  \label{39}
\end{equation}
Therefore $\Sigma _{s}$ is the constant of the quark mass renomalization due
to the presence of the quark-antiquark pairing. In the particular case of
the vanishing bare quark mass,

\[
M=0, 
\]
the system of equations (35) becomes

\begin{eqnarray}
\Sigma ^{\left( -\right) } &=&\Sigma ^{\left( +\right) }gK_{\Sigma ^{\left(
+\right) }}^{0}\left( 0,\beta ,\mu \right) ,  \nonumber \\
\Sigma ^{\left( +\right) } &=&\Sigma ^{\left( -\right) }gK_{\Sigma ^{\left(
-\right) }}^{0}\left( 0,\beta ,\mu \right) .  \label{40}
\end{eqnarray}
It follows that

\begin{equation}
\Sigma ^{\left( +\right) }=\pm \Sigma ^{\left( -\right) }.  \label{41}
\end{equation}
If

\[
\Sigma ^{\left( +\right) }=\Sigma ^{\left( -\right) } 
\]
then

\[
\Sigma _{t}=0 
\]
and $\Sigma _{s}$ is determined by the equation

\begin{equation}
gK_{\Sigma _{s}}^{0}\left( 0,\beta ,\mu \right) =1.  \label{42}
\end{equation}
If

\begin{equation}
\Sigma ^{\left( +\right) }=-\Sigma ^{\left( -\right) }  \label{43}
\end{equation}
then 
\begin{equation}
\Sigma _{s}=0  \label{44}
\end{equation}
and $\Sigma _{t}$ is determined by the equation

\begin{equation}
gK_{\Sigma _{t}}^{0}\left( 0,\beta ,\mu \right) =-1.  \label{45}
\end{equation}
The existence of the solution of either the equation (42) or the equation
(45) depends on the sign of the constant $g$: For $g>0$ the equation (45)
has no solution, while in the case $g<0$ the solution of the equation (42)
does not exist.

The existence of a non-vanishing solution of either the equation(42) or the
equation (45) at the appropriate values of the coupling constant $g$ would
mean the spontaneous breaking of the chiral invariance in the system of the
quarks and antiquarks with the vanishing bare quark mass: the
quark-antiquark pairing makes the massless quarks to become the massive
ones. In the case of the non-vanishing solution $\Sigma _{t}$ of the
equation (45) the flavor symmetry is spontaneously broken while in the case
of the non-vanishing solution $\Sigma _{s}$ of the equation (42) there is no
spontaneous breaking of the flavor symmetry.

\noindent \underline{Phase 2.} In this phase from the equations (25) and
(26) we derive following system of equations for the constants $\Sigma
^{\left( \pm \right) }$ and $\Pi _{s}:$

\begin{eqnarray}
\Sigma ^{\left( +\right) } &=&\left[ M+\Sigma ^{\left( -\right) }\right]
gK_{\Sigma ^{\left( -\right) }}^{\Pi _{s}}\left( M,\beta ,\mu \right) , 
\nonumber \\
\Sigma ^{\left( -\right) } &=&\left[ M+\Sigma ^{\left( +\right) }\right]
gK_{\Sigma ^{\left( +\right) }}^{\Pi _{s}}\left( M,\beta ,\mu \right) ,
\label{46}
\end{eqnarray}
and 
\begin{equation}
\Pi _{s}=-\Pi _{s}\frac{1}{2}g\left\{ K_{\Sigma ^{\left( +\right) }}^{\Pi
_{s}}\left( M,\beta ,\mu \right) +K_{\Sigma ^{\left( -\right) }}^{\Pi
_{s}}\left( M,\beta ,\mu \right) \right\} ,  \label{47}
\end{equation}
Because

\[
\Pi _{s}\neq 0, 
\]
from the equation (47) it follows that

\begin{equation}
\frac{1}{2}g\left\{ K_{\Sigma ^{\left( +\right) }}^{\Pi _{s}}\left( M,\beta
,\mu \right) +K_{\Sigma ^{\left( -\right) }}^{\Pi _{s}}\left( M,\beta ,\mu
\right) \right\} =-1.  \label{48}
\end{equation}
The solution of this equation may exist only if $g$ is negative. If the bare
quark mass does not vanish,

\[
M\neq 0, 
\]
then from the equation (46) it follows that

\[
\Sigma ^{\left( +\right) }=\Sigma ^{\left( -\right) } 
\]
and therefore

\[
\Sigma _{t}=0. 
\]
The flavor symmetry is not spontaneously broken, and the equation (48)
becomes

\begin{equation}
gK_{\Sigma _{s}}^{\Pi _{s}}\left( M,\beta ,\mu \right) =-1,  \label{49}
\end{equation}
From the equations (46) now we obtain

\begin{equation}
\Sigma _{s}=-\frac{M}{2},  \label{50}
\end{equation}
and the order parameter $\Pi _{s}$ is determined by the equation

\begin{equation}
gK_{0}^{\Pi _{s}}\left( \frac{M}{2},\beta ,\mu \right) =-1.  \label{51}
\end{equation}
The existence of $\Sigma _{s}$ and $\Pi _{s}$  would mean the quark
mass renormalization and the spontaneous parity violation without the
spontaneous breaking of the flavor symmetry. \hspace{0in}\noindent If the
bare quark mass is vanishing,

\[
M=0, 
\]
then the system of equations (46) becomes

\begin{eqnarray}
\Sigma ^{\left( +\right) } &=&\Sigma ^{\left( -\right) }gK_{\Sigma ^{\left(
-\right) }}^{\Pi _{s}}\left( 0,\beta ,\mu \right) ,  \nonumber \\
\Sigma ^{\left( -\right) } &=&\Sigma ^{\left( +\right) }gK_{\Sigma ^{\left(
+\right) }}^{\Pi _{s}}\left( 0,\beta ,\mu \right) .  \label{52}
\end{eqnarray}
From these equations we derive the relation

\begin{equation}
gK_{\Sigma ^{\left( -\right) }}^{\Pi _{s}}\left( 0,\beta ,\mu \right)
.gK_{\Sigma ^{\left( +\right) }}^{\Pi _{s}}\left( 0,\beta ,\mu \right) =1.
\label{53}
\end{equation}
The system of two equations (48) and (53) for two variables $gK_{\Sigma
^{\left( -\right) }}^{\Pi _{s}}\left( 0,\beta ,\mu \right) $ and\linebreak $%
gK_{\Sigma ^{\left( +\right) }}^{\Pi _{s}}\left( 0,\beta ,\mu \right) $ has
the unique solution

\begin{equation}
gK_{\Sigma ^{\left( -\right) }}^{\Pi _{s}}\left( 0,\beta ,\mu \right)
=gK_{\Sigma ^{\left( +\right) }}^{\Pi _{s}}\left( 0,\beta ,\mu \right) =-1.
\label{54}
\end{equation}
Substituting this value of $gK_{\Sigma ^{\left( \pm \right) }}^{\Pi
_{s}}\left( 0,\beta ,\mu \right) $ into the equations (52), we obtain the
relation

\begin{equation}
\Sigma ^{\left( +\right) }=-\Sigma ^{\left( -\right) },  \label{55}
\end{equation}
which means that

\begin{equation}
\Sigma _{s}=0.  \label{56}
\end{equation}
The equation (54) becomes

\begin{equation}
gK_{\Sigma _{t}}^{\Pi _{s}}\left( 0,\beta ,\mu \right) =-1.  \label{57}
\end{equation}
Denote $\Sigma _{\max }$ the solution of the equation

\begin{equation}
gK_{\Sigma _{\max }}^{0}\left( 0,\beta ,\mu \right) =-1.  \label{58}
\end{equation}
A comparison of two equations (57) and (58) gives the relation

\begin{equation}
\Sigma _{t}^{2}+\Pi _{s}^{2}=\Sigma _{\max }^{2}.  \label{59}
\end{equation}
Thus in the case of the massless bare quarks with the negative coupling
constant $g$ if the equation (58) has a solution $\Sigma _{\max }\neq 0$
then there exists a continuous class of the quark-antiquark pair condensates
spontaneously breaking the flavor symmetry and the parity conservation with
the order parameters $\Sigma _{t}$ and $\Pi _{s}$ satisfying the condition
(59).

\noindent \underline{Phase 3}. In this phase from the equations (25) and
(26) we derive the system of equations for the constants $\Sigma _{s}$ and $%
\Pi ^{\left( \pm \right) }$:

\begin{equation}
\Sigma _{s}=\left( M+\Sigma _{s}\right) \frac{1}{2}g\left\{ K_{\Sigma
_{s}}^{\Pi ^{(-)}}\left( M,\beta ,\mu \right) +K_{\Sigma _{s}}^{\Pi ^{\left(
+\right) }}\left( M,\beta ,\mu \right) \right\} ,  \label{60}
\end{equation}

\begin{eqnarray}
\Pi ^{\left( +\right) } &=&-\Pi ^{(-)}gK_{\Sigma _{s}}^{\Pi ^{(-)}}\left(
M,\beta ,\mu \right) ,  \nonumber \\
\Pi ^{\left( -\right) } &=&-\Pi ^{(+)}gK_{\Sigma _{s}}^{\Pi ^{(+)}}\left(
M,\beta ,\mu \right) .  \label{61}
\end{eqnarray}
From the equations (61) it follow that

\begin{equation}
\Pi ^{\left( +\right) }=\pm \Pi ^{\left( -\right) }.  \label{62}
\end{equation}
Because

\[
\Pi _{t}\neq 0, 
\]
we must choose the solution with

\begin{equation}
\Pi ^{\left( +\right) }=-\Pi ^{\left( -\right) }  \label{63}
\end{equation}
and therefore

\begin{equation}
\Pi _{s}=0,  \label{64}
\end{equation}
Then the equations (61) give

\begin{equation}
gK_{\Sigma _{s}}^{\Pi _{t}}\left( M,\beta ,\mu \right) =1.  \label{65}
\end{equation}
From the equations (60) and (65) it follows that 
\[
M=0. 
\]
Thus the phase 3 may exist only in the case of the vanishing bare quark
mass. In this case $\Sigma _{s}$ and $\Pi _{t}$ are determined by the
equation

\begin{equation}
gK_{\Sigma _{s}}^{\Pi _{t}}\left( 0,\beta ,\mu \right) =1  \label{66}
\end{equation}
Denote $\Pi _{\max }$ the solution of the equation

\begin{equation}
gK_{0}^{\Pi _{\max }}\left( 0,\beta ,\mu \right) =1.  \label{67}
\end{equation}
We have also the relation of the form (59) between the parameters $\Sigma
_{s},$ $\Pi _{t}$ and $\Pi _{\max }:$

\begin{equation}
\Sigma _{s}^{2}+\Pi _{t}^{2}=\Pi _{\max }^{2}.  \label{68}
\end{equation}
Thus in the case of the massless bare quarks with the positive coupling
constant $g$ if the equation (67) has a solution $\Pi _{\max }\neq 0$, then
there exists a continuous class of the quark-antiquark pair condensates
spontaneously breaking the flavor symmetry and the parity conservation with
the order parameters $\Sigma _{s}$ and $\Pi _{t}$ satisfying the condition
(68).

\section{DISCUSSIONS.}

\qquad We have shown that in a system of quarks and antiquarks with two
flavors and $N_{c}$ co-lors and with the direct four-fermion coupling
determined by the interaction Lagrangian (3) and the coupling constant
matrix (24) the spontaneous breaking of the parity conservation and/or the
flavor symmetry without that of the color symmetry may take place only if
one among the equations (51), (58) and (67) has some non-vanishing solution
which would determines the order parameters of the corresponding
spontaneously breaking symmetry phase. There is a significant difference
between these equations for the quark-antiquark pairing and the BCS equation
for the superconductivity: According to the formula (28) for the function $%
K_{\Sigma }^{\Pi }\left( M,\beta ,\mu \right) $the expression under the
integration over the momentum in the new equations (51), (58) and (67) is
finite, while in the BCS equation the expression under the integration over
the momentum is divergent at the Fermi surface. Therefore the solution of
the BCS equation always exists, while the equation (51), (58) and (67) may
have the solutions only if the magnitude of the coupling constant $g$ is
large enough and the range of the form-factor $\varphi \left( p\right) $ is
wide enough. To simplify the calculations in the sequel we replace the
form-factor $\varphi \left( p\right) $ by introducing the momentum cut-off $%
\lambda .$This means that we set

\[
\varphi \left( p\right) =\theta \left( \lambda -p\right) . 
\]
Then the equation (28) becomes

\begin{eqnarray}
K_{\Sigma }^{\Pi }\left( M,\beta ,\mu \right) &=&\frac{1}{8\pi ^{2}}%
\int_{0}^{\lambda }\frac{1}{E_{\Sigma }^{\Pi }\left( p,M\right) }\left\{ 
\mathrm{th}\frac{\beta \left[ E_{\Sigma }^{\Pi }\left( p,M\right) +\mu \right] 
}{2}\right.  \nonumber \\
&&+\left. \mathrm{th}\frac{\beta \left[ E_{\Sigma }^{\Pi }\left( p,M\right)
-\mu \right] }{2}\right\} p^{2}dp,  \label{69}
\end{eqnarray}

We discuss first the case of the massive bare quarks 
\[
M\neq 0. 
\]
The existence of some non-vanishing solution $\Pi _{s}$ of the equation (51)
would mean the spontaneous violation of the parity conservation without the
flavor symmetry breaking. It is easy to verify that $\Pi _{s}\neq 0$ does
exist if the coupling constant $g$ and the momentum cut-off $\lambda $
satisfy following (sufficient) condition

\begin{equation}
\left| g\right| \frac{\lambda ^{2}}{8\pi ^{2}}f\left( \frac{M}{2\lambda }%
\right) >1  \label{70}
\end{equation}
with

\begin{equation}
f\left( \alpha \right) =\sqrt{1+\alpha ^{2}}-\alpha ^{2}\ln \left( \frac{1}{%
\alpha }+\sqrt{1+\frac{1}{\alpha ^{2}}}\right) .  \label{71}
\end{equation}
For the free energy density $F_{\beta }^{M}\left[ \Pi _{s}\right] $ in this
phase we have the relation 
\begin{eqnarray}
\beta F_{\beta }^{M}\left[ \Pi _{s}\right] &=&4\left\{ \frac{1}{2}\left( -%
\frac{M^{2}}{4}+\Pi _{s}^{2}\right) K_{0}^{\Pi _{s}}\left( \frac{M}{2},\beta
,\mu \right) \right.  \nonumber \\
&&\left. -\int_{0}^{1}\left[ -\frac{M^{2}}{4}\left( 2-\omega \right) +\omega
\Pi _{s}^{2}\right] K_{0}^{\omega \Pi _{s}}\left( M\left( 1-\frac{\omega }{2}%
\right) ,\beta ,\mu \right) d\omega \right\} .  \label{72}
\end{eqnarray}
The spontaneously violating parity phase is a stable one if and only if

\begin{eqnarray}
&&\frac{1}{2}\left( -\frac{M^{2}}{4}+\Pi _{s}^{2}\right) K_{0}^{\Pi
_{s}}\left( \frac{M}{2},\beta ,\mu \right)  \nonumber \\
&&\qquad \leq \int_{0}^{1}\left[ -\frac{M^{2}}{4}\left( 2-\omega \right)
+\omega \Pi _{s}^{2}\right] K_{0}^{\omega \Pi _{s}}\left( M\left( 1-\frac{%
\omega }{2}\right) ,\beta ,\mu \right) d\omega
\end{eqnarray}
For the given values of $\beta ,\mu $ and $\lambda $ this condition is
satisfied only for the large enough values of the ratios $\Pi _{s}/M$

\begin{equation}
\frac{\Pi _{s}}{M}\geq C\left( \beta ,\mu ,\frac{\lambda }{M}\right) .
\label{74}
\end{equation}
The curve $C\left( \infty ,0,x\right) $ as a function of $x=\lambda /M$ is
plotted in Fig.1.
\newpage
\ \ 
% Syntax:  \centereps{<width>}{<height>}{<path+filename>}
%\centereps{9.8365cm}{6.0671cm}{fig1.eps}
\centerline{\includegraphics[width=98mm]{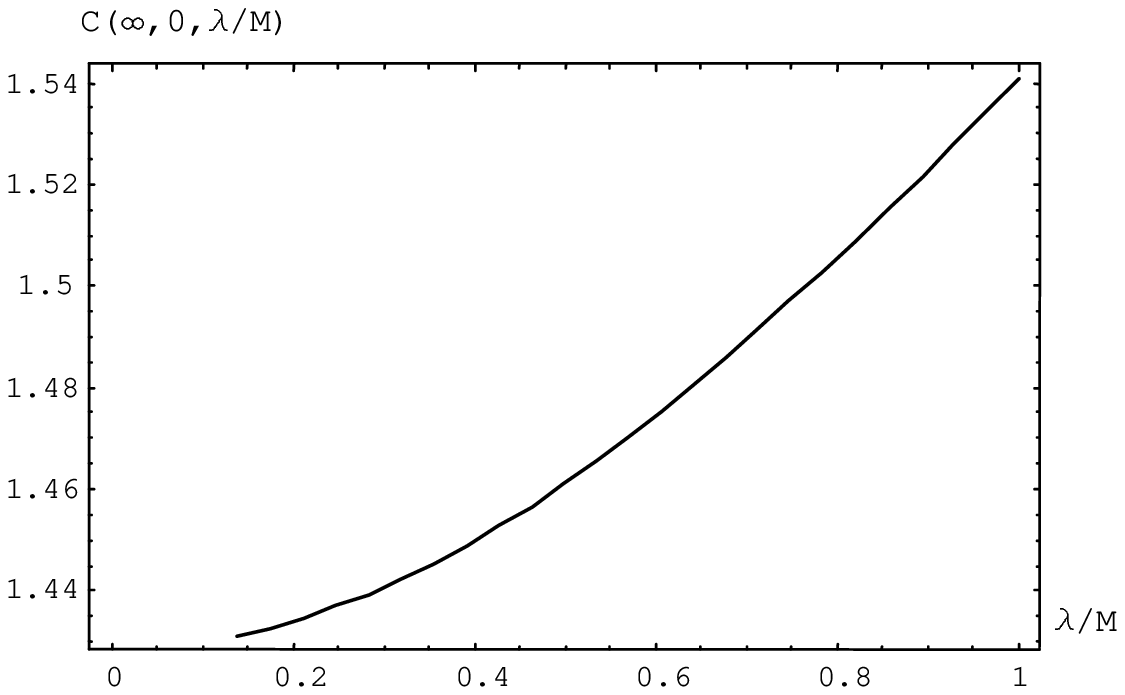}}
\begin{center}
{\small {\bf Fig. 1.} The curve $C\left( \infty ,0,\frac{\lambda }{M}%
\right) $ as a function of $x=\lambda /M.$}
\end{center}
Now we discuss the case of the massless bare quarks 
\[
M=0. 
\]
If the coupling constant $g$ and the momentum cut-off $\lambda $ satisfy the
(sufficient) condition 
\begin{equation}
\frac{1}{8\pi ^{2}}\left| g\right| \lambda ^{2}>1,  \label{75}
\end{equation}
then the equation (57) with negative $g$ and the equation (66) with positive 
$g$ have non-vanishing solutions $\Sigma _{t}$, $\Pi _{s}$ (phases 2) and $%
\Sigma _{s}$, $\Pi _{t}$ (phases 3), respectively. In these phases the
parity conservation and /or the flavor $SU\left( 2\right) $ symmetry are
spontaneously broken. Their free energy densities are determined by
following relations:

\noindent \underline{Phase 1}

\begin{equation}
\beta F_{\beta }^{0}\left[ \Sigma \right] =4\Sigma ^{2}\left\{ \frac{1}{2}%
K_{\Sigma }^{0}\left( 0,\beta ,\mu \right) -\int_{0}^{1}K_{\omega \Sigma
}^{0}\left( 0,\beta ,\mu \right) d\omega \right\} .  \label{76}
\end{equation}

\noindent \underline{Phases 2}

\begin{equation}
\beta F_{\beta }\left[ \Sigma _{t},\,\Pi _{s}\right] =4\left( \Sigma
_{t}^{2}+\Pi _{s}^{2}\right) \left\{ \frac{1}{2}K_{\Sigma _{t}}^{\Pi
_{s}}\left( 0,\beta ,\mu \right) -\int_{0}^{1}\omega K_{\omega \Sigma
_{t}}^{\omega \Pi _{s}}\left( 0,\beta ,\mu \right) d\omega \right\} .
\label{77}
\end{equation}
\underline{Phases 3} 
\begin{equation}
\beta F_{\beta }\left[ \Sigma _{s},\,\Pi _{t}\right] =4\left( \Sigma
_{s}^{2}+\Pi _{t}^{2}\right) \left\{ \frac{1}{2}K_{\Sigma _{s}}^{\Pi
_{t}}\left( 0,\beta ,\mu \right) -\int_{0}^{1}\omega K_{\omega \Sigma
_{s}}^{\omega \Pi _{t}}\left( 0,\beta ,\mu \right) d\omega \right\} .
\label{78}
\end{equation}

Because 
\begin{equation}
K_{\Sigma _{t}}^{\Pi _{s}}\left( 0,\beta ,\mu \right) =K_{\Sigma _{\max
}}^{0}\left( 0,\beta ,\mu \right) ,  \label{79}
\end{equation}
$\Sigma _{\max }$ being determined by the relation (59), and 
\begin{equation}
K_{\Sigma _{s}}^{\Pi _{t}}\left( 0,\beta ,\mu \right) =K_{0}^{\Pi _{\max
}}\left( 0,\beta ,\mu \right) ,  \label{80}
\end{equation}
$\Pi _{\max }$ being determined by the relation (68), the free energy
densities of all phases 2 with different pairs of order parameters $\Sigma
_{t}$, $\Pi _{s}$ have one and the same value 
\begin{equation}
F_{\beta }\left[ \Sigma _{t},\,\Pi _{s}\right] =F_{\beta }\left[ \Sigma
_{\max },0\right] =F_{\beta }^{0}\left[ \Sigma _{\max }\right] ,  \label{81}
\end{equation}
and those of all phases 3 with different pairs of order parameters $\Sigma
_{s}$, $\Pi _{t}$ also have one and the same value 
\begin{equation}
F_{\beta }\left[ \Sigma _{s},\,\Pi _{t}\right] =F_{\beta }\left[ 0,\Pi
_{\max }\right] =F_{\beta }^{0}\left[ \Pi _{\max }\right] .  \label{82}
\end{equation}

\noindent Thus we have the total degeneracy between different phases in each
continuous class. The free energy densities (77) and (78) are negative, and
the phases 2 and 3 are stable. Note that in these phases not only the parity
conservation and/or the $SU\left( 2\right) $ flavor symmetry, but also the
chiral symmetry, are spontaneously broken. Phase 1 is a limiting case of
phases 3.

The expressions (31) and (32) of the order parameters $\Sigma _{\left(
b_{j}\right) }^{\left( a_{i}\right) }$ and $\Pi _{\left( b_{j}\right)
}^{\left( a_{i}\right) }$ contain the arbitrary unit vector $\mathbf{n}$.
However the free energy densities (77) and (78) do not depend on the
direction of this vector.

The critical temperature of the phase transition is determined by the
equation 
\begin{equation}
\left| g\right| K_{0}^{0}\left( 0,\beta _{c},\mu \right) =1.  \label{83}
\end{equation}
The solution $\beta _{c}$ at given values of $\left| g\right| $, $\lambda $
(satisfying the condition (75)) and $\mu $ is plotted in Fig. 2. At each
temperature lower than the critical one, $\beta >\beta _{c}$, the order
parameter $\Sigma _{\max }$ or $\Pi _{\max }$ is non-vanishing. The $\beta $%
-dependence of $\Sigma _{\max }$ or $\Pi _{\max }$ at some given values of $%
\left| g\right| $, $\lambda $ (satisfying the condition (75)) and $\mu $ is
presented in Fig. 3.

\centerline{\includegraphics[width=80mm]{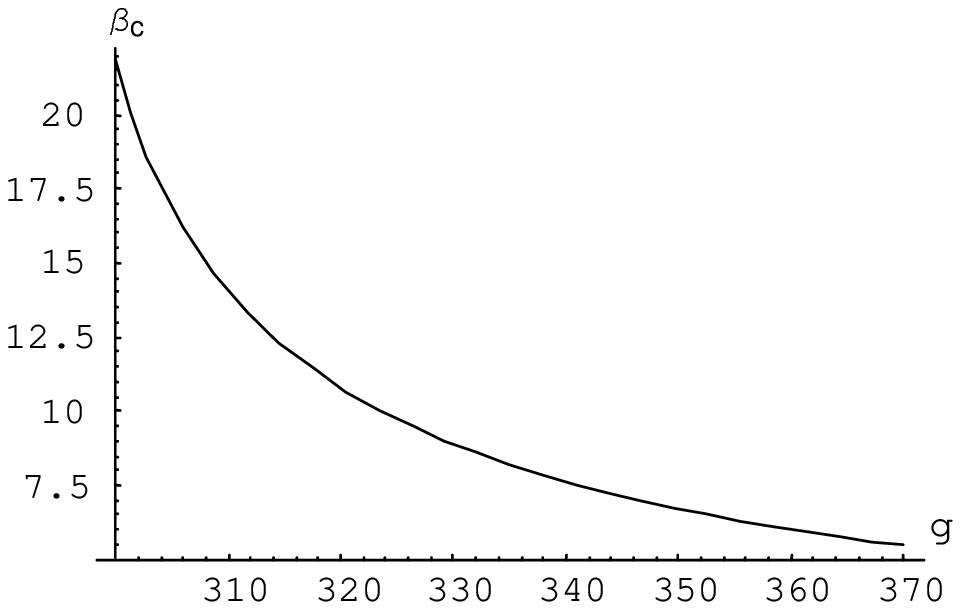}}
%\centereps{8cm}{5cm}{fig2.eps}
\begin{center}
{\small {\bf Fig. 2.} $\beta _{c}$ in $\left( GeV\right)^{-1}$ vs. $\left| g\right| $ in $\left(
GeV\right) ^{-2}$ at $\lambda = 0.6GeV$ and $\mu =0.3GeV.$}
\end{center}
\centerline{\includegraphics[width=90mm]{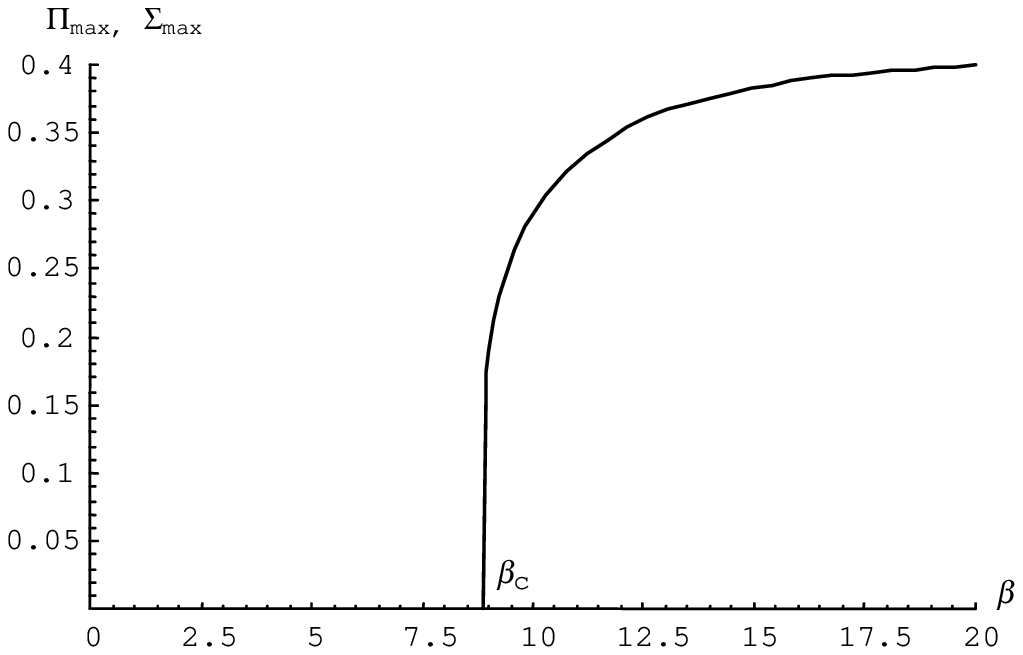}}
%\centereps{9cm}{6cm}{fig3.eps}
\begin{center}
{\small {\bf Fig. 3.} $\Pi _{\max }$ or $\Sigma_{\max }$ in $GeV$ vs. $\beta $ in $\left( GeV\right) 
^{-1}$ at $g=330\left( GeV\right)^{-2}$,\\ $\lambda =0.6GeV,$ $\mu =0.3GeV.$}
\end{center}
\newpage
\begin{center}
{\bf APPENDIX.}
\end{center}

In the following tables we present the values of the function $%
K_{0}^{0}\left( M,\beta ,\mu \right) $ at given values of the constants $M$, 
$\lambda $, $\mu $ in the $GeV$ unit and $\beta $ in the $\left( GeV\right)
^{-1}$unit

\begin{center}
{\small {\bf Table 1.} $K_{0}^{0}\left( M,\beta ,\mu\right) $ at $M=0.1$ and $\lambda =0.6$.}
\centerline{\includegraphics[width=6in]{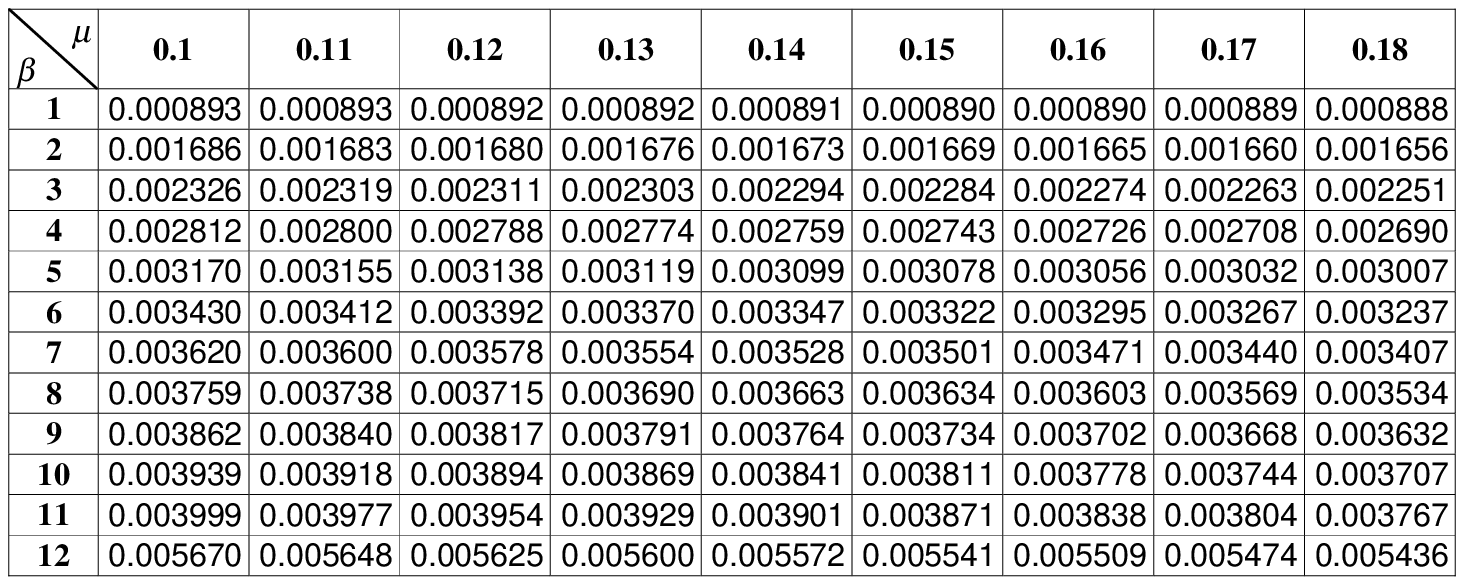}}
%\centereps{6.1436in}{2.4in}{tab1.eps}
{\small {\bf Table 2.} $K_{0}^{0}\left( M,\beta ,\mu \right) $ at $M=0.1$ and $\lambda =0.7$.}
\centerline{\includegraphics[width=6in]{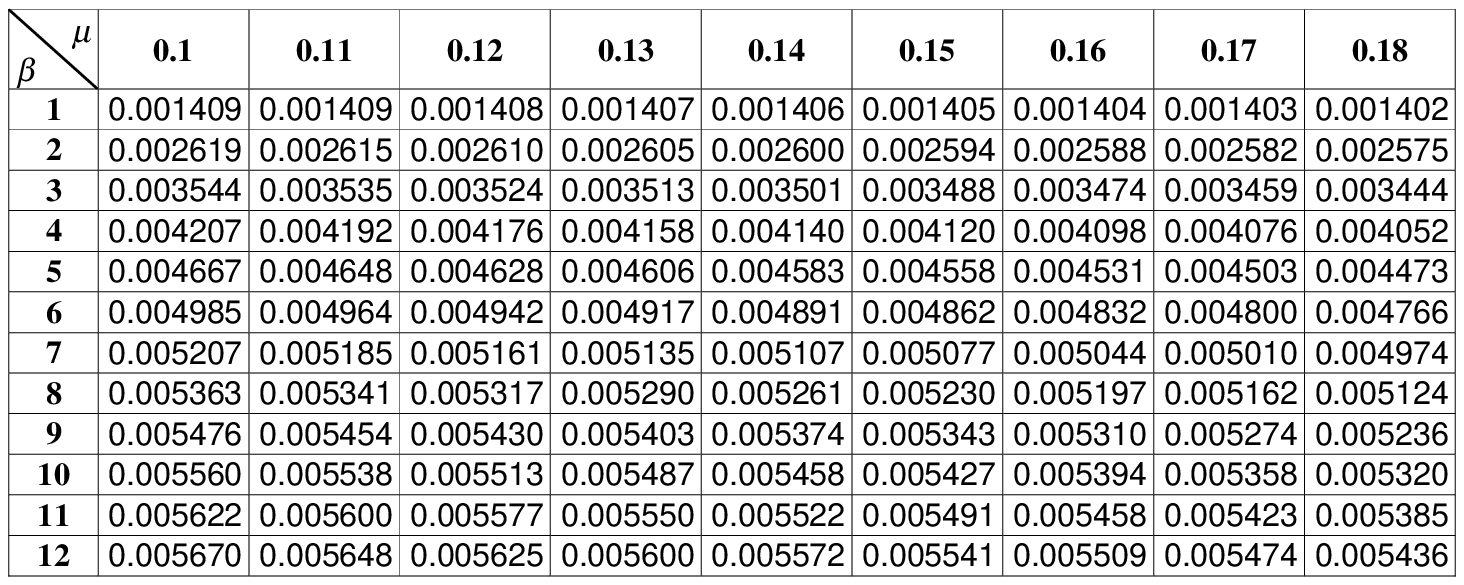}}
%\centereps{6.1618in}{2.4in}{tab2.eps}
{\small {\bf Table 3.} $K_{0}^{0}\left( M,\beta ,\mu\right) $ at $M=0.1$ and $\lambda =0.8$.}
\centerline{\includegraphics[width=6in]{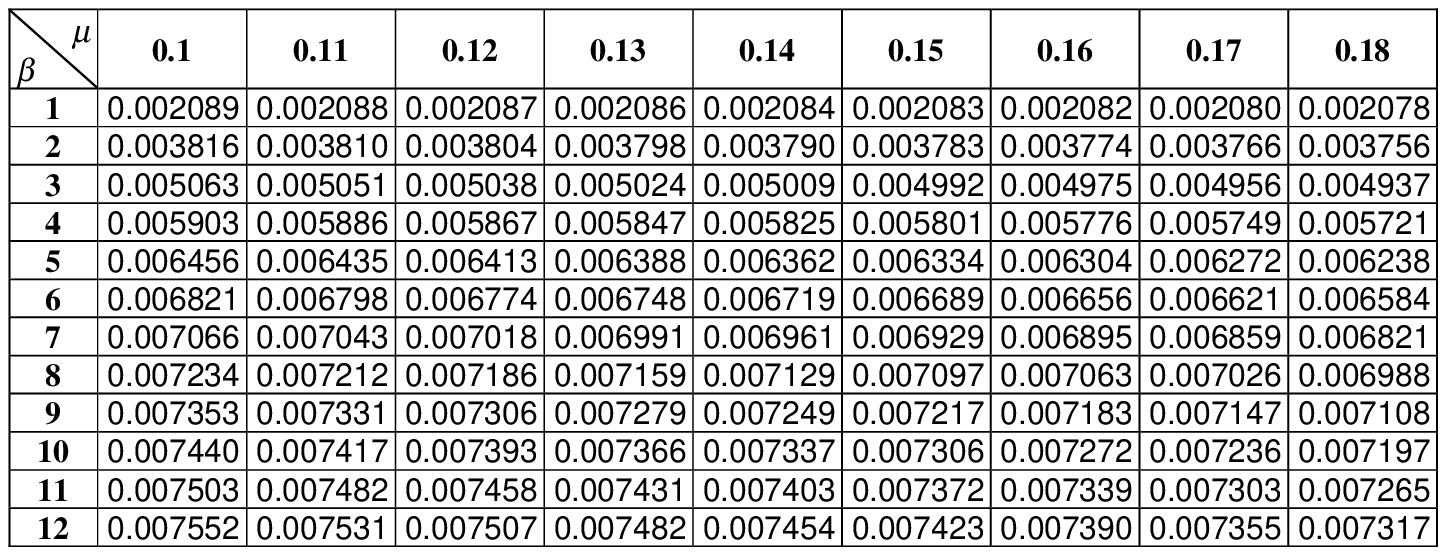}}
%\centereps{6.103in}{2.4in}{tab3.eps}
\end{center}
\newpage
\begin{center}
{\small {\bf Table 4.} $K_{0}^{0}\left( M,\beta ,\mu \right) $ at $M=0.13$ and $\lambda=0.8$.}
\centerline{\includegraphics[width=6in]{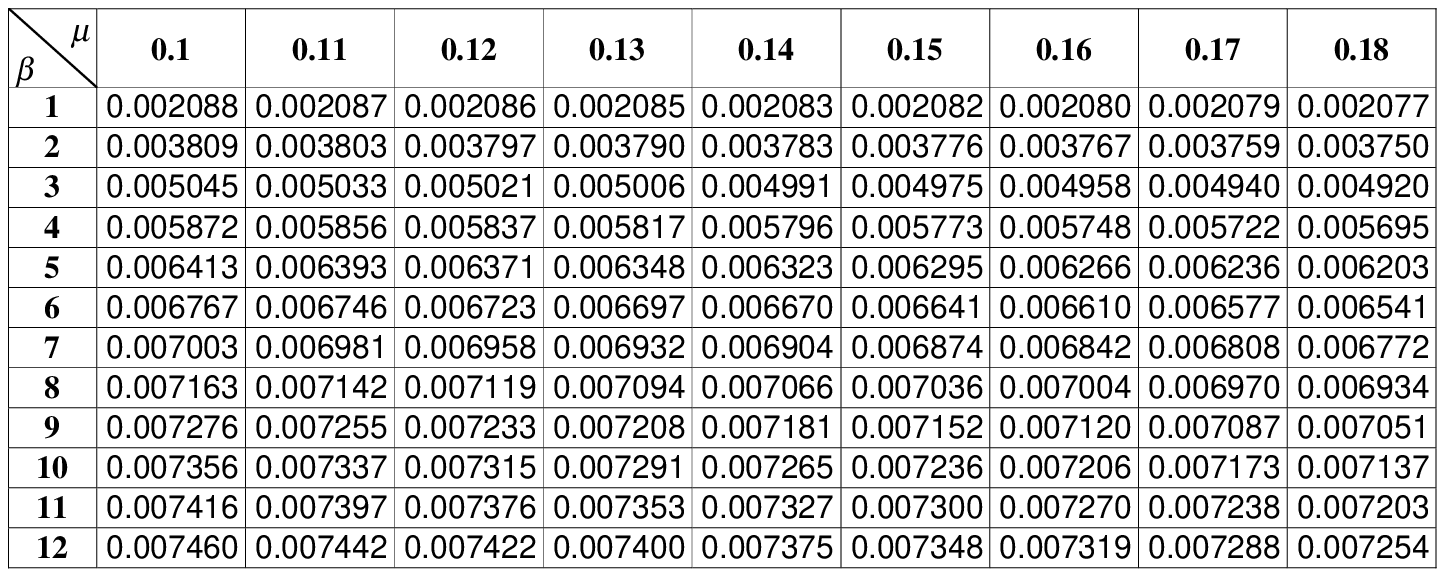}}
%\centereps{5.8963in}{2.5071 in}{tab4.eps}
{\small {\bf Table 5 .} $K_{0}^{0}\left( M,\beta ,\mu\right) $ at $M=0.15$ and $\lambda =0.8$.}
\centerline{\includegraphics[width=6in]{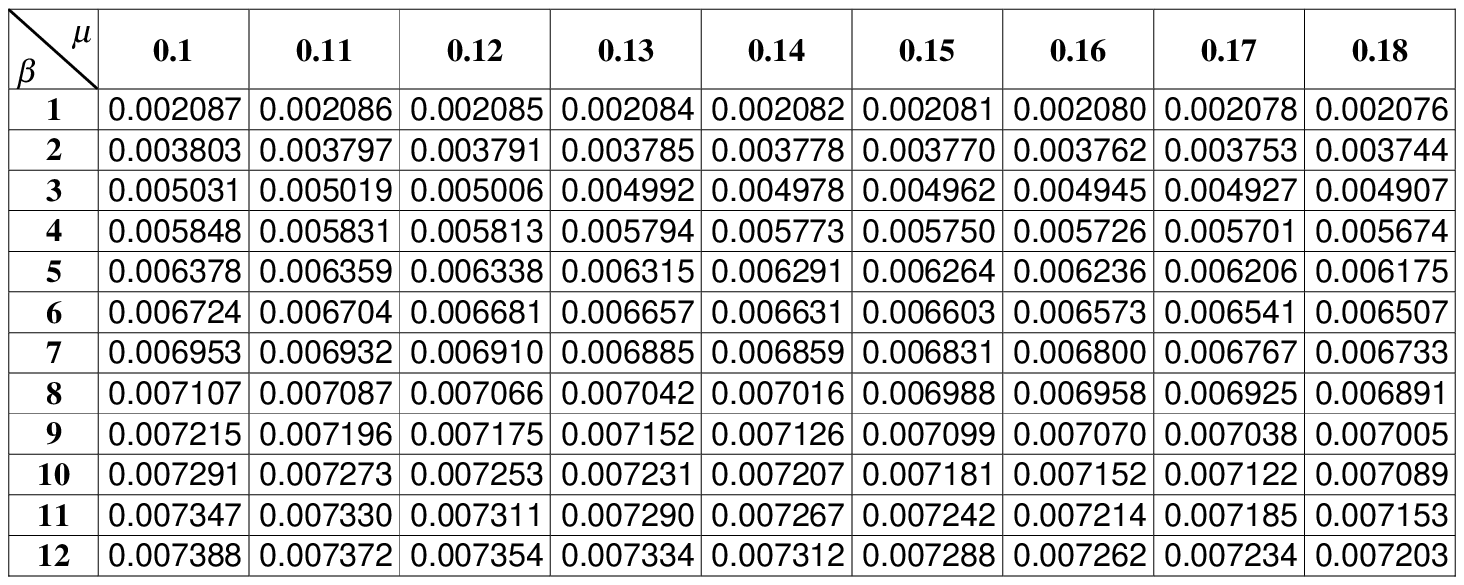}}
%\centereps{5.8072in}{2.5062 in}{tab5.eps}
\end{center}
\vskip 0.2cm
\noindent{\bf ACKNOWLEDGMENT.}
\vskip 0.1cm
The authors express their sincere thank to the Natural Sciences
Council for the financial support to this work.\newline

\end{document}